# Plasmonic Hot Electron Transport Driven Site-Specific Surface-Chemistry with Nanoscale Spatial Resolution

Emiliano Cortés[1], Wei Xie[2], Javier Cambiasso[1], Adam S. Jermyn[3,4], Ravishankar Sundararaman[4,5],

Prineha Narang[4,6,7], Sebastian Schlücker[2], Stefan A. Maier[1]

[1] The Blackett Laboratory, Department of Physics, Imperial College London, London SW7 2AZ, UK.
[2] Physical Chemistry I, Department of Chemistry and Center for Nanointegration Duisburg-Essen (CENIDE), University of Duisburg-Essen, Universitätsstr. 5, Essen 45141, Germany.
[3] Institute of Astronomy, Cambridge University, Cambridge CB3 0HA, UK.
[4] Joint Center for Artificial Photosynthesis, California Institute of Technology, 1200 East California Boulevard, Pasadena, California 91125, USA.
[5] Department of Materials Science and Engineering, Rensselaer Polytechnic Institute, 110 8th street, Troy, New York 12180, USA.
[6] Thomas J. Watson Labs of Applied Physics, California Institute of Technology, 1200 East California Boulevard, Pasadena, California 91125, USA.
[7] NG NEXT, 1 Space Park Drive, Redondo Beach, California 90278, USA.

E.C. and W.X. contributed equally.
Correspondence address: E.C. (e.cortes@imperial.ac.uk), P.N. (prineha@caltech.edu) and S.S. (sebastian.schluecker@uni-due.de)

**Abstract:** Nanoscale localization of electromagnetic fields near metallic nanostructures underpins the fundamentals and applications of plasmonics.[1,2] The unavoidable energy loss from plasmon decay, initially seen as a detriment, has now expanded the scope of plasmonic applications to exploit the generated hot carriers.[3,4] However, quantitative understanding of the spatial localization of these hot carriers, akin to electromagnetic near-field maps, has been elusive. Here we spatially map hot-electron-driven reduction chemistry with 15 nanometre resolution as a function of time and electromagnetic field polarization for different plasmonic nanostructures. We combine experiments employing a six-electron photo-recycling process that modify the terminal group of a self-assembled monolayer on plasmonic silver nanoantennas, with theoretical predictions from first-principles calculations of non-equilibrium hot-carrier transport in these systems. The resulting localization of reactive regions, determined by hot carrier transport from high-field regions, paves the way for hot-carrier extraction science and nanoscale regio-selective surface chemistry.

**Introduction:** Nanostructured materials that present plasmonic resonances enable intense light focusing, mediating electromagnetic (EM) energy transfer from the far to the near field or vice versa. Thus, these nanostructures can be considered as optical nanoantennas and are key elements in the conversion of free-space light to evanescently confined modes in nanometre-scale volumes below the diffraction limit.[1,5] For metals such as Au and Ag, localized surface plasmon resonances (LSPRs) in nanoantennas fall into the optical regime. Due to the sub-wavelength character of these modes, the electric energy density is significantly higher than the magnetic counterpart. Self-sustaining electromagnetic oscillations then require an additional energy term, found in the form of a kinetic energy density of the free carriers of the metal.[6, 7] Sub-diffraction electric field concentration at visible wavelengths in metals is only possible due to the existence of these energetic carriers, highlighting the mixed light-matter mode nature of LSPRs.

Light incident on metallic nanoantennas excites surface plasmons, which may decay to hot carriers through several mechanisms including direct interband transitions, phonon-assisted intraband transitions and geometry-assisted transitions (collectively referred to as Landau damping).[6, 8-10] In recent years, increased attention has been paid to these loss mechanisms, leading to the extension of the concept of plasmonic nanoantennas not only as sub-diffraction light-focusing objects, but also as reactive elements in the interplay between light and nanoscale materials. The possibility to extract and use hot carriers generated after the non-radiative decay of surface plasmons (SPs) has triggered new research within the field. The unavoidable drawback of losses at visible wavelengths in metals has now turned into an exciting opportunity for energy conversion, photodetection, photochemistry, and photocatalysis.[3, 4, 11, 12]

In spite of being very energetic, these hot carriers cannot travel over distances larger than 10s of nanometres before they lose their energy via scattering.[13] After being generated, the excitations will decay on timescales ranging from a few tens of femtoseconds to picoseconds via a series of ultrafast processes such as electron-electron scattering and equilibration to the lattice by electron-phonon scattering.[8, 14, 15] These scattering events thermalize the carriers and bring their energies closer to the Fermi level of the metal, on average. Plasmonic hot carrier applications, on the other hand, require carriers far from the Fermi level to more efficiently drive reactions in molecules and to overcome activation barriers. Therefore, to describe hot carriers that impinge on the surface of a nanostructure for collection requires a full description of the hot carrier transport, including the spatial and temporal evolution of energy and momentum.[4, 9, 16] Typically transport of excited carriers is described using one of two regimes: i) a ballistic regime where carriers rarely scatter within the structure and ii) a diffusive regime where several scattering events can occur within the characteristic structure dimensions. Hot carrier transport falls between the ballistic and diffusive regimes – an intermediate regime where the plasmonic structure is neither much smaller nor significantly larger than a mean free-path between sequential scattering events. Far-from-equilibrium hot carriers are a particular challenge for current Monte Carlo or Boltzmann transport methods.[17]

The carriers generated by plasmon decay impinge upon the surface of a plasmonic nanostructure to be collected, either directly or after scattering against other carriers and phonons in the metal. Collection (extraction) of hot carriers is currently accomplished via two primary methods: Schottky barrier-based devices or molecular species able to capture the electrons and/or holes from the interface. The former needs a semiconductor at the interface of the metallic nanostructure or antenna capable of accepting the carriers into its conduction band; this method has led to important developments in applications such as photodetection [18, 19], nanoimaging[20], and artificial photosynthesis[21], among others. The latter instead involves injection of hot carriers from the metal structure into available orbitals of nearby adsorbates.[22] Essentially the proposed mechanism for plasmon-driven chemistry involves the injection of an electron from the metal into an anti-bonding state of an adsorbed molecule, causing either desorption or the dissociation of a bond in the adsorbate.[12, 23] Depending on the potential energy landscape of these transient ions they can further react on the metal surface[24] or diffuse and react later on in solution.[25] Hot-electron mediated reactions at Al, Ag and Au plasmonic interfaces have been achieved by using these ideas, enabling such encouraging work as $O_2$ and $H_2$ dissociation, among others.[26-28]

Despite this significant progress in our understanding of hot-carrier generation, relaxation dynamics and extraction, there are still many open questions that need to be addressed. Among them, the possibility of mapping the reactivity of plasmonic antennas with nanometre resolution is critical, as it would guide the efficient design and fabrication of reactive nanoantennas for plasmon-induced energy conversion or photocatalysis, for example.[29] While theoretical and experimental capabilities developed in nanophotonics to accurately predict and measure the electric near-field distribution in plasmonic antennas have enabled countless applications that accelerated the field of plasmonics to its current mature state[1, 2], a thorough understanding in the context of interfacial hot-carrier distributions is still lacking. The local reactivity of these systems is dictated by a combination of hot-carrier generation, distribution and extraction, and therefore multiple processes need to be taken into account in the experimental design and theoretical modelling.

In this report we present experimental and theoretical results of hot-electron driven reactivity mapping with nanometre resolution. We use the ability of hot electrons to locally reduce the terminal group of a self-assembled molecular layer that covers the surface of a nanoantenna. This modification is spatially localized using 15 nm nanoparticles (NPs) specially designed to react with the converted molecules only. SEM imaging is used to record the position of the reporter-NPs after different time periods of light illumination and hence plasmon excitation, thus progressively mapping the reactivity of the nanoantennas. Geometry-dependent differences have been observed between Ag bow-ties and rounded Ag dimer antennas. We give physical insight into the experimentally observed spatial reactivity distributions with the aid of an efficient computational technique that we developed to predict hot carrier transport with spatial and energy resolution in complex plasmonic nanostructures from first principles. This analysis is essential to understand the efficiency of plasmonic energy conversion and plasmonic hot carrier extraction using molecular species. The transport method developed here for non-equilibrium carriers is an important tool in the prediction and explanation of this complex dynamical process. Using this method, we present quantitative theoretical predictions for hot carrier transport in Ag bowties that are in agreement with the experimental results. The combination of top-down and bottom-up fabrication approaches, selective surface chemistry and theoretical calculations of the transport of non-equilibrium carriers has allowed us to show that nanoscale regions with high electromagnetic fields in high-curvature areas are the most reactive locations within plasmonic antennas. Hot-electron driven nanoscale patterning of surface chemistry in plasmonic antennas can now be employed for self-guided positioning of catalytic materials or molecules to highly reactive regions (reactive spots) and for the selective modification of high EM field regions (hot spots).

## Results:

**Modifying the local surface chemistry in Ag antennas by hot-electron mediated reactions.**

Let us begin by describing the experimental approach to trace reactivity in plasmonic antennas (Figure 1). Briefly, Ag nanoantennas on quartz substrates were fabricated by electron-beam lithography. Their spectral features, near-field distribution and exact dimensions are described in later sections. Once fabricated and characterized, Ag nanoantennas were modified with 4-nitrothiophenol (4-NTP). As a thiolated molecule, 4-NTP forms a densely packed monolayer on the surface of the Ag antennas (Figure 1a). When 4-NTP coated Ag antennas are illuminated at their plasmon resonance frequency in the presence of an acid halide media (i.e. HCl, HBr, HI), they undergo a six-electron-mediated reduction to form 4-aminothiophenol (4-ATP).[30] Key elements for driving this reaction are hot electrons, protons and halide anions. Hot electrons are generated within silver and transferred to molecules adsorbed on the metal surface, as shown by wavelength-dependent studies (see Figure S1). Protons serve as the hydrogen source. Halide ions are required for the photorecycling of electron-donating Ag atoms. Briefly, hot holes generated on the Ag surface - upon hot-electron extraction - are filled by halide ions (i.e. Cl$^-$). The insoluble silver halide present on the Ag surface (i.e. AgCl) undergoes a photodissociation reaction that regenerates the Ag surface. A series of control experiments demonstrate that without this photorecycling counter half-reaction, the six-hot-electron reduction cannot proceed.[30]

By selecting proper conversion times, this reaction should locally change the chemical composition of the antennas only at highly reactive regions, whilst most of the antenna would remain covered by the original (unconverted) molecules (Figure 1b). However, extracting spatially resolved chemical information at the nanoscale from a surface by a non-invasive method is not a straightforward task. Our system is sensitive to light, therefore tools like tip-enhanced Raman spectroscopy (TERS) should be avoided for mapping reactivity.[31] Using surface enhanced Raman scattering (SERS) for demonstrating the chemical conversion is also not conclusively meaningful, since only molecules in the electromagnetic hot spots can be detected due to dominance of the electromagnetic field enhancement.[32] In our approach we used 15 nm AuNPs functionalized with 11-mercaptoundecanoic acid (MUA) and the well-known 1-Ethyl-3-(3-dimethylaminopropyl)-carbodiimide/ N-hydroxysuccinimide (EDC/NHS) reaction (Figure 1c) to report on the hot-electron conversion (see Supplementary Materials for details). This highly specific and high-yield coupling reaction between the terminal carboxylic acid group in the AuNPs and the terminal amino (converted) groups in the Ag antennas is used to form an amide bond[33] that serves as a reporter of the local hot-electron reactivity on the antennas (Figure 1d). High-resolution SEM imaging is finally performed to track the position of the AuNPs in the Ag antennas. The fact that the NPs were added after illumination (under dark conditions) and that they were guided only by the local chemical reactivity of the antennas allowed us to avoid high EM field trapping effects that could have biased the results.[34, 35] Indeed, tracking reactive regions in plasmonic antennas during illumination should be avoided, as high EM fields can have significant effects on the local concentration of reactants, masking hot-carrier reactive spots with EM field hot spots. Deposition of conducting reporters such as polymers or metal ions can generate an additional channel for electron migration once nucleation has started, also biasing the results.[36-38] Finally, we analysed SEM images of the Au particles attached to 100 Ag antennas, illuminated for different lengths of time with different incident polarizations and geometries, to reveal the local hot-electron reactivity in these systems.

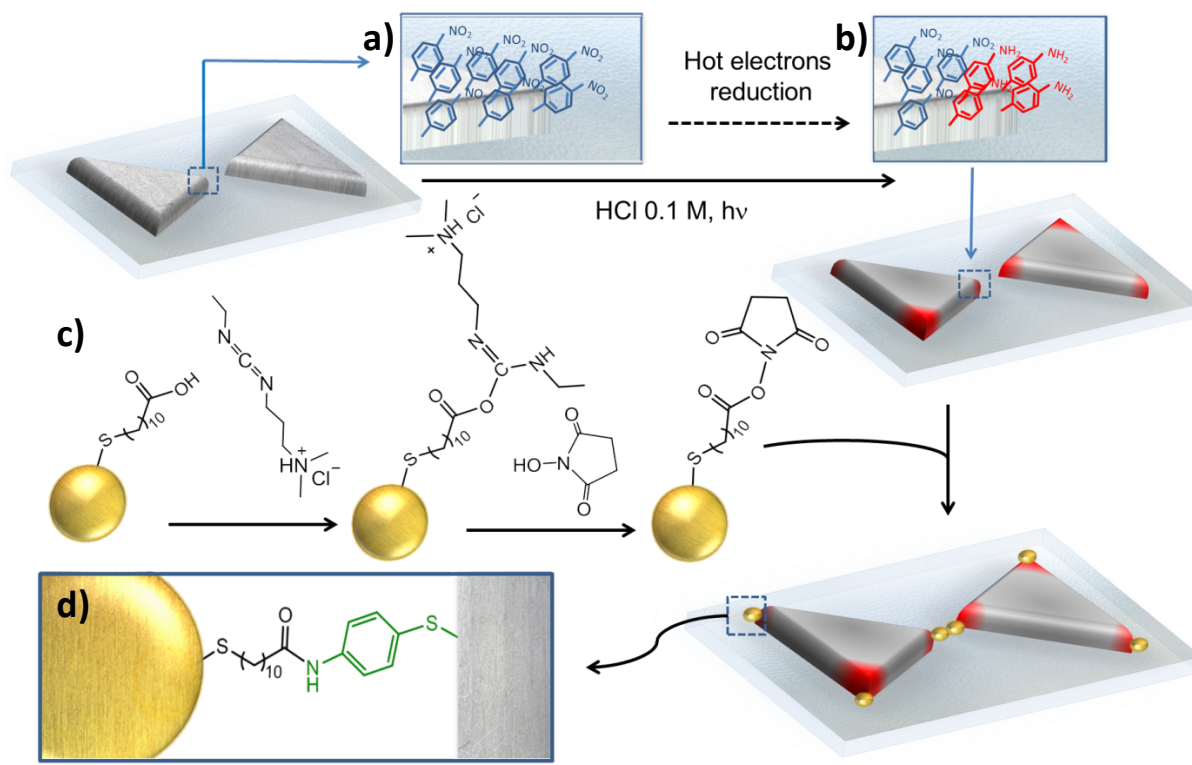

**Figure 1| Scheme of the local surface chemistry modification and AuNPs tracking approach. a)** Ag nanoantennas were modified overnight with 1 mM ethanolic solution of 4-nitrothiophenol (4-NTP). Several ethanol/water washing steps were performed on each sample. **b)** 4-NTP coated antennas were immersed in 0.1 M HCl solution and illuminated for different times at their LSPR wavelength (633 nm) with a power density of 1 W/cm$^2$. Samples were rinsed with water and immediately dipped in the activated AuNP suspension. c) 15 nm AuNPs coated with 11-mercaptoundecanoic acid (MUA) as a capping layer were suspended in HEPES buffer and mixed with 1 mM 1-Ethyl-3-(3-dimethylaminopropyl)-carbodiimide (EDC) and 1 mM N-hydroxysuccinimide (NHS) and left to react for 30 minutes followed by 2 purification centrifugation steps. d) The activated and purified AuNPs were left in contact with the hot-electron converted Ag antennas to react overnight, thus creating the amide (-NH-C=O) bond. Several washing steps with HEPS buffer and water were performed prior to 2 nm Pt sputtered coating for SEM imaging.

We fabricated two different types of antennas: Ag bow-ties (BT) and Ag bar-dimers (BD). Finite-difference time-domain (FDTD) simulations were conducted to determine the range of Ag antenna sizes that exhibit a plasmon resonance at 633 nm in water (see Supplementary Materials for simulation details). Ag antennas were then fabricated by electron-beam lithography (see Supplementary Materials for details on fabrication and characterization). Each sample is composed of an array of 100 antennas of the same dimensions. Ag markers (1 µm in size) were added at each corner of the array and served as non-plasmonic control surfaces. Figures 2 shows the FDTD simulated near-field electric field distribution for the obtained dimensions of Ag-BT (Ag-BD) antennas illuminated at 633 nm in water for parallel - Figure 2a and (2d) - and perpendicular – Figure 2b and (2e) - polarizations. Colour scale bars represent the electric field enhancement ($|E|/|E_0|$) values obtained in each case. We also show the simulated scattering, absorption and extinction spectra for each case. Figure 2c (2f) shows typical SEM images for the obtained Ag-BT (Ag-BD) antennas and also single-antenna scattering spectra measured in air for each case. In this way, we have arrays of antennas with different shapes and tuned spectral characteristics to probe the localization-conversion experiments. We have measured single-antenna scattering spectra over a number of antennas on each array, confirming a highly uniform spectral response (see Figure S2).

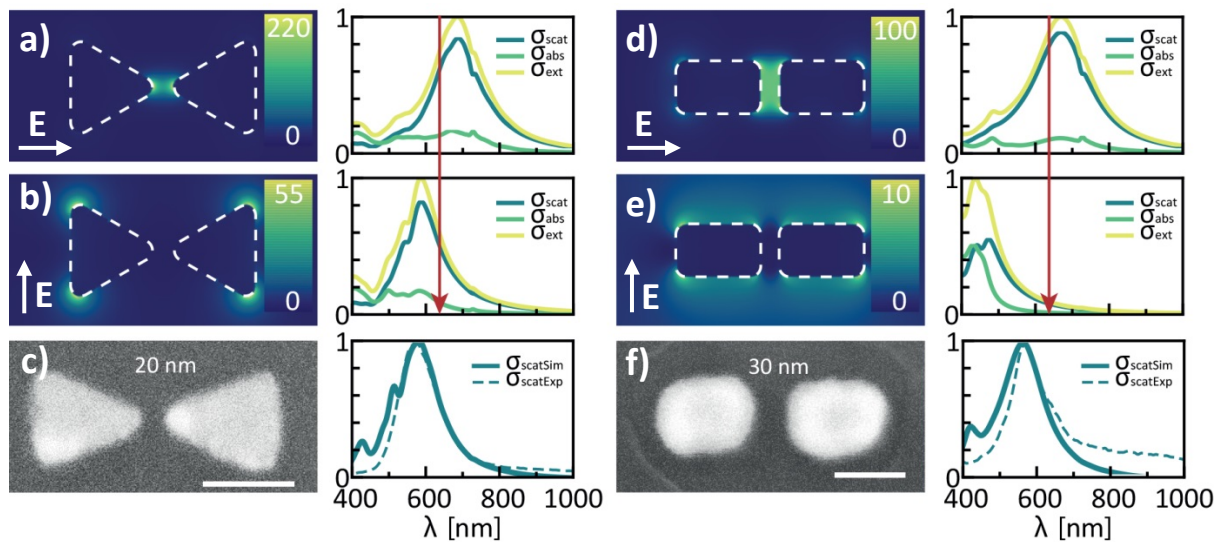

**Figure 2| Plasmonic response of Ag bow-ties and Ag bar-dimer antennas. a-b)** FDTD simulations of the near-field distribution for a Ag bow-tie antenna at 633 nm in water for parallel (a) and perpendicular (b) polarized illumination. Colour scale bars represent the field enhancement ($|E|/|E_0|$) values obtained in each case. Simulated scattering (blue), absorption (green) and extinction (yellow) spectra for these polarizations are shown to the right. Red arrow highlights the laser wavelength used in the conversion experiments (633 nm). **c)** SEM image of a typical Ag bow-tie antenna (gap 20 nm) and simulated (full line) and measured (dotted line) single-antenna scattering spectra in air. Scale bar: 80 nm. **d-e)** FDTD simulations of the near-field distribution for a Ag bar-dimer antenna at 633 nm in water for parallel (d) and perpendicular (e) polarized illumination. Colour scale bars represent the field enhancement ($|E|/|E_0|$) values obtained in each case. Simulated scattering (blue), absorption (green) and extinction (yellow) spectra. Red arrow highlights the laser wavelength used in the conversion experiments (633 nm). **f)** SEM image of a typical Ag bar-dimer antenna (gap 30nm) and simulated (full line) and measured (dotted line) single-antenna scattering spectra in air. Scale bar: 100 nm.

### Experimental mapping of hot-electron conversion in Ag antennas.

Once the antennas were fabricated and characterized, we performed the conversion experiments as described in Figure 1. However a series of control experiments were necessary first to show that conversion takes place at the single antenna level, and also that the proposed surface chemistry reactions were efficient and specific enough to trace the hot-electron converted molecules.

In order to show that conversion takes place under our experimental conditions, we performed single-antenna surface-enhanced Raman spectroscopy (SERS) to detect the conversion from 4-NTP to 4-ATP (Figure 3a). High EM field confinement in our antennas (Figure 2) allowed label-free monitoring of the reaction by in-situ SERS spectroscopy. By employing 4-NTP coated Ag antennas in the presence of 0.1 M

HCl and by using a 10 µW diffraction-limited spot at λ=633 nm (i.e. at the plasmon resonance of our antennas), the C–C and C–S stretching bands of 4-ATP appeared at ~1,590 and ~1,080 cm$^{-1}$, respectively (Figure 3a). Temporal spectral series showed that 4-NTP Raman peaks decreased while 4-ATP peaks increased. Raman cross-sections (σ) for both molecules are not the same ($σ_{4-NTP} > σ_{4-ATP}$), so it is not possible to make a linear correlation of this behaviour; however they followed the expected trend – shifting the laser excitation wavelength away from $λ_{max}$ of the plasmon peak leads to lower reduction activity (see Figure S1). These results confirm that the six-hot-electron reduction reaction can proceed in our system to completely convert 4-NTP into 4-ATP.

After confirming the conversion reaction in our antennas, we moved towards testing the surface chemistry reactions. A negative control experiment was performed by incubating a 4-NTP modified Ag antenna sample with the activated 15 nm AuNPs (activation proceeds as described in Figure 1 by EDC/NHS chemistry). SEM imaging of the antennas and surrounding Ag films (markers) showed that, as expected, no reporter particles remain attached on either the Ag antennas, the Ag film or the quartz surface (Figure 3b). By contrast, when the same experiment was performed under exactly the same conditions on a 4-ATP coated sample, we detected 15 nm AuNPs attached to both the Ag film and the Ag antennas with no trace of the particles on the quartz substrate (Figure 3c). These two control experiments indicate that the reaction is highly specific to the presence of 4-ATP and that the amide-bond formation proceeds at a very high yield under these experimental conditions (inset Figure 3c). Other less effective reactions and conditions were also tested (see Supplementary Materials). Finally, we performed a control experiment on a partially-converted sample: a 4-NTP functionalized Ag-antenna array was illuminated at 633 nm with a 0.5 mm spot at 1 W for 2 minutes in 0.1 M HCl and 15 nm activated AuNPs were added right after finishing the illumination. The markers and the entire array were therefore illuminated at the same time. As shown in Figure 3d, some NPs appeared to attach only to the antenna (i.e. no particles on the Ag film or on the quartz substrate). This last control shows that LSPRs are necessary for the reaction to take place.[39] The probability of interband optical absorption is very low for Ag at this excitation wavelength (633 nm) and therefore chemical conversion mediated by this absorption method should not occur on the Ag film;[40] only carriers derived from the non-radiative decay of LSPRs should contribute to the reduction of 4-NTP. These experiments demonstrate that the method described in Figure 1 is suitable to monitor the local changes in reactivity of the plasmonic antennas by hot-electron mediated reduction chemistry.

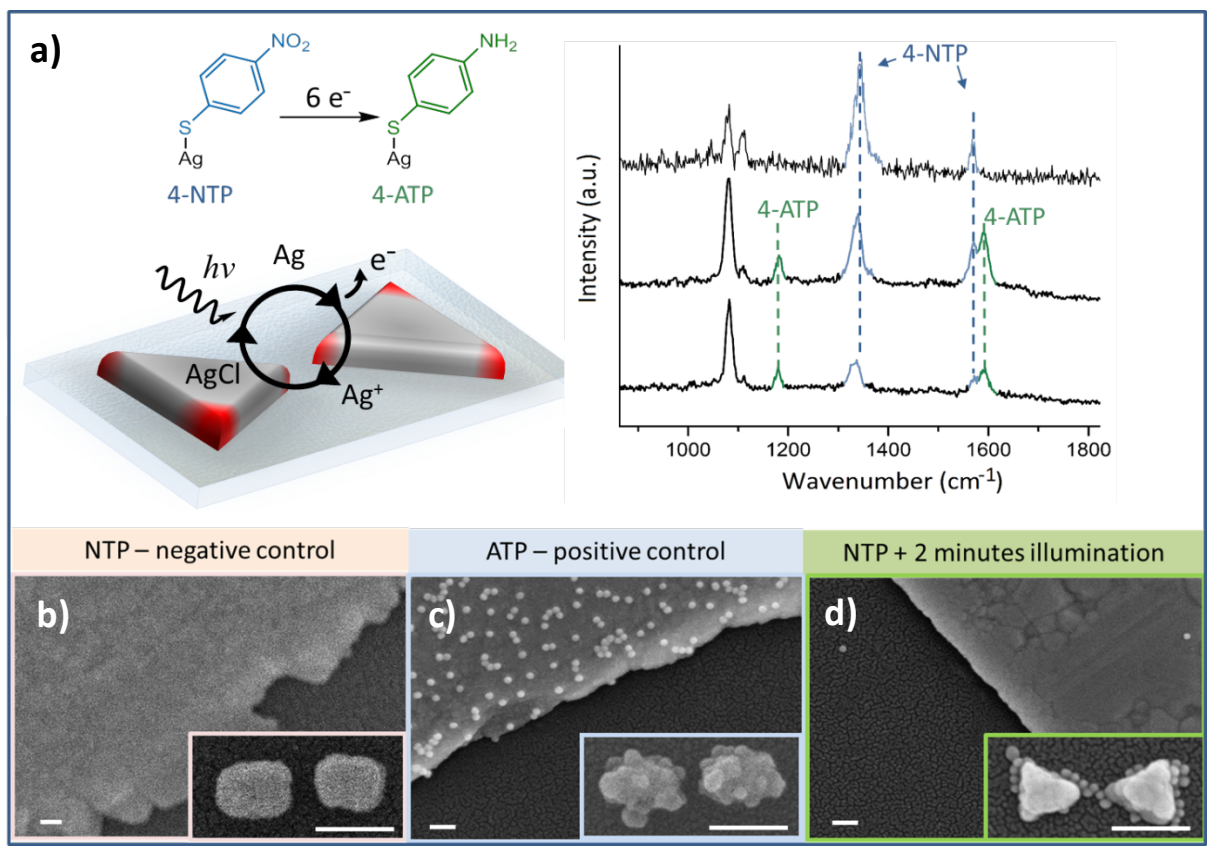

**Figure 3|** In-situ label-free SERS monitoring of hot-electron mediated reduction and control experiments. **a)** Single-antenna SERS detection of hot-electron reduction from 4-NTP to 4-ATP in the presence of 0.1 M HCl, λ = 633 nm, power 1 mW, integration time 5 sec. Time-dependent spectra highlighting the conversion from 4-NTP (top spectrum) to 4-ATP (bottom spectra). **b)** SEM images of the negative control experiment: sample was coated with 4-NTP and left to react with the activated 15 nm AuNPs. No particles were detected on either the Ag film or on the Ag antennas (inset). **c)** SEM images of the positive control experiment: sample was coated with 4-ATP and leave to react with the activated 15 nm AuNPs. After several washes of the sample with HEPES buffer and water, the amide reaction was detected both on the Ag film and the Ag antennas (inset). **d)** SEM images of the partial-conversion experiment: Ag antennas were coated with 4-NTP and illuminated at 633 nm for 2 minutes in 0.1 M HCl at a power density of 1 mW/cm$^2$. AuNPs were found attached only to the Ag-antennas whilst no particles were found on the substrate or in the Ag film. All the samples were treated under exactly the same conditions, followed by the same washing steps prior to 2-3 nm Pt coating for SEM imaging. Scale bars: 100 nm for all images.

### Transport, dynamics and spatial dependence of hot carrier distributions in Ag bow-ties

In order to obtain physically meaningful results, the experimental methodology described above relies on a highly site-specific extraction of energetic carriers from the nanoantennas. We investigate this assumption now theoretically via first principles. The spatial dependence of hot-carrier driven reduction reactions depends on the spatial distribution of sufficiently energetic hot electrons reaching the surface.

This distribution depends both on the initial spatial and energy distributions of hot carriers generated by plasmon decay, and the subsequent transport of the generated electrons to the surface.

Our theoretical predictions start with the initial spatial and energy distribution of hot carriers generated by plasmon decay,

$$P_0(E,\mathbf{r}) = \mathrm{Im}\varepsilon(\omega,E)\,|\mathbf{E}(\mathbf{r})|^2/(2\pi\hbar), \qquad (1)$$

where $\mathrm{Im}\varepsilon(\omega,E)$ is the *ab initio* frequency-dependent imaginary dielectric function, resolved by the carrier energy $E$ generated by the responsible electronic transition, and $\mathbf{E}(\mathbf{r})$ is the electric field distribution inside the metal upon illumination obtained from electromagnetic simulations. The calculated dielectric function includes contributions due to direct interband transitions[10] and phonon-assisted intraband transitions[14] in general; although only the latter contribute to carrier generation in silver at 633 nm illumination (≈2 eV photons, below the interband threshold ≈3.6 eV).

We then calculate the carrier flux $\phi_n(E,\mathbf{r})$ reaching the surface after $n$ scattering events as

$$\phi_n(E,\mathbf{r}) = \int d\mathbf{r}'\, P_n(E,\mathbf{r}')\, e^{-|\mathbf{R}|/\lambda(E)}\, \mathbf{R}\cdot\mathbf{n}/\,(4\pi|\mathbf{R}|^3) \text{ and} \qquad (2)$$

$$P_n(E,\mathbf{r}) = \int d\mathbf{r}' \int dE'\, P(E|E') P_{n-1}(E',\mathbf{r}')\, e^{-|\mathbf{R}|/\lambda(E')}/\,(4\pi|\mathbf{R}|^2\lambda(E')). \qquad (3)$$

Above, $P_n(E,\mathbf{r})$ is the distribution of hot carriers generated after $n$ scattering events, $\mathbf{R} := \mathbf{r} - \mathbf{r}'$ is the separation between source and target points, and $\mathbf{n}$ is the unit surface normal vector. We use *ab initio* calculations of electron-electron and electron-phonon scattering to determine the energy-dependent carrier mean-free path $\lambda(E)$ and the probability $P(E|E')$ that the scattering of a carrier of energy $E'$ generates a carrier of energy $E$. We assume that the number of generated hot carriers is negligible compared to the number of thermal electrons in the metal, such that these quantities depend only on the carrier energy and the background distribution. The final hot carrier flux $\phi(E,\mathbf{r}) = \sum_n \phi_n(E,\mathbf{r})$, accounts for carriers that reach the surface without scattering, and after an arbitrary number of scattering events. This approach efficiently solves a linearized Boltzmann equation with an *ab initio* collision integral[8], assuming that each scattering event randomizes the carrier momentum, which is an excellent approximation for carriers with $|E-E_f| \ll E_f$, the Fermi energy.[41] See the Supplementary Materials for further details.

Figure 4 shows the predicted hot carrier flux reaching the surface of an Ag bowtie antenna. We show the spatially-resolved probabilities of carriers with energy E greater than a threshold $E_{cut}$, which can for example be interpreted as the minimum carrier energy required to drive a chemical reaction. The highest hot carrier flux is at the tip of the bowtie antenna where the field is strongest and then decays exponentially away from the tip. The exponential decay length depends both on the plasmon field distribution inside the metal (the skin-depth length scale) and the mean-free path of the hot carriers. Hot electrons of lower energy reach further from the field hot spots due to their longer mean-free path and because they can be generated after multiple scattering events, while higher energy electrons remain localized closer to the field hot spots. The high $E_{cut}$ necessary to generate a sufficient density of extracted carriers to drive the six-electron reduction reaction results in the high spatial resolution of the chemistry observed in the experiments as follows.

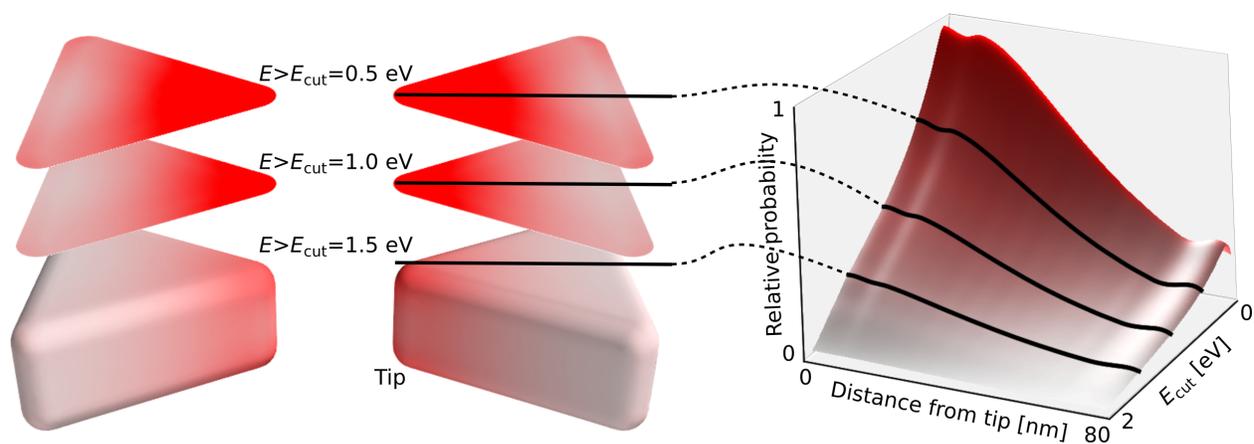

**Figure 4| First-principles predictions of spatial and energy-resolved probabilities of plasmonic hot carriers that reach the surface of a Ag bow-tie antenna under 633 nm illumination.** These predictions are based on *ab initio* calculations of hot carrier generation including phonon-assisted intraband excitations, and a transport model that accounts for multiple scattering events and the energy-dependent electron mean-free path from *ab initio* calulations. The left panel shows the relative flux of hot electrons with energy greater than a threshold $E_{cut}$ (relative to the Fermi level of silver) on the surface of a bow-tie antenna illuminated at resonance. The right panel shows the corresponding distribution as a function of $E_{cut}$, averaged over planes of constant distance from the tip. The probability drops linearly with increasing $E_{cut}$, and exponentially with distance from the tip due to the plasmon field distribution inside the metal and the transport of hot carriers from the point of generation to the surface. Higher $E_{cut}$ results in lower hot carrier flux, but greater spatial resolution.

Figure 5 presents the results of our efforts to localize hot-electron mediated reactions. We evaluated geometry, time, and polarization dependence on hot-electron transfer to 4-NTP in Ag antennas. Figure 5a shows the results for parallel illumination (λ = 633 nm) in Ag bow-ties after 1 minute of conversion (in 0.1 M HCl). Representative SEM images clearly point towards the fact that the hot-spot tips of the structure are the most reactive regions (i.e. they exhibit preferential binding of the Au reporter-NPs). A map containing the localization of reporter-NPs from all 100 antennas in the array is presented at the bottom of the column (see Supplementary Materials and Figure S3 for further details). We repeat that the 15 nm AuNP reporters are added after the conversion (with no illumination), thus they are only guided by the surface chemistry reactivity. This fact allows us to avoid optical trapping and high EM field effects that could potentially mask the results. Our observations are in line with previous and current theoretical considerations on carrier extraction from metallic nanoantennas[9]. Photoemission spectroscopy of essentially free electrons from plasmonic antennas[42] has shown preferential emission at field hot spot regions; here, we demonstrate similar site-selective extraction of bound carriers to locally drive surface chemistry, though we emphasize we are not collecting photo-emitted electrons. Once carriers have been generated, the possibility that they reach the surface and escape (i.e. transfer to the adsorbed 4-NTP molecule) before the onset of significant energy loss occurs inside the material is strongly dependent on the curvature of the structure. This explains the strong reactivity observed near the tips, consistent with our calculations presented in Figure 4. These results are in line with the observation of Moskovits et al[36], as the randomization of the momenta of hot electrons occurs on a shorter time scale, with the first scattering event, than the time for complete electron thermalization; therefore the geometry of the antenna becomes the crucial parameter for hot carriers to reach the surface. Nanoscale confinement of carriers also plays an important role for their final extraction capabilities, with importance of subtler aspects such as defects and/or crystal orientation within the metal nanoantenna. We would expect an anisotropic distribution in the localization of the reporter-AuNPs if carrier collection was dominated by defects in our polycrystalline structures. However, clear geometry-dependent reactive spots for short illumination times were detected in our experiments (Figure 5a bottom panel). Hot electron travel distances to their final extraction points are greatly reduced in our nanoantennas compared to recent results shown for a 3 μm semiconductor wire, where defects and impurities dominate the extraction of carriers exited over the band-gap.[29] For this illumination time, where we have few particles per antenna we have noticed a strong dependence in the localization map regarding the orientation of the antenna with respect to the illumination polarization (see Figure S3). Antennas presenting tilts over 20° (respect to the main axes) present also Au NPs reporters at the corners of the antenna, see the second SEM image in Figure 5a. These results are in line with our theoretical predictions and polarization-dependent field distribution (Figure 2a and 2b). These conclusions are further supported by the following additional observations.

Figure 5b shows the localized particles in Ag bow-ties after 2 minutes of illumination with parallel polarization. The progression of the reaction can be followed and we note that after the central tips, the edges and corners seem to be the most reactive regions within this geometry. For a carrier generated inside the material, the probability of reaching the surface is higher at the edges/corners in comparison to the flat-topped surface due to curvature. For Ag bow-ties, our results imply then that reactivity is highest at sharp tips and lowest on flat planar sections of the structure. In particular, this requires feature sizes that are smaller than the hot carrier mean-free path, an energy-dependent quantity that results in ≈ 10 nm for electrons 2 eV above the Fermi level in silver.[14] . This indicates that in our six-electron reaction the spatial confinement is governed by a combination of the density of carriers that reach the surface in a given time period and the energy cut-off. As shown, for longer illumination times we can detect particles far from where they are primarily generated from which we infer that very low-energy carriers are able to convert 4-NTP to 4-ATP. Our calculations point towards that spatial confinement should also be achievable with reactions involving higher energetic barriers. Then the number of electrons involved or the energy-barrier of the reaction should act as spatial filters for hot-carriers driven process. However, the site-specific spatial confinement can also be lost if the final acceptor of the electrons (coating the antennas) delocalize them, as recently shown for polyvinylpyrrolidone-induced anisotropic growth of Au nanoprisms in plasmon-driven synthesis.[38]

To evaluate the geometry-dependent reactivity in Figure 5c, we show representative SEM images for Ag bar-dimers after 1 minute of parallel illumination. For these rounded structures without sharp edges or corners the localization maps (Figure 5c, bottom panel) differ dramatically from those observed for Ag bow-ties. In this case, the reporter particles are always found on top of the antennas, with no preference for the surrounding regions. Based on these results, if we imagine two nanostructures with the same absorption cross section ($\sigma_{Abs}$) at a given resonant wavelength ($\lambda$) but shaped in different geometries, the one with sharper features would produce hot carriers that should be more efficiently extracted before their energy is lost. Finally, with perpendicularly polarized illumination, we could not detect any attached particles over the entire array (Figure 5d). This is in line with our findings with the Ag film control experiment (Figure 3d), demonstrating that non-radiative plasmon decay is the main source of hot electrons driving the reaction. As expected, absorption in these structures for this polarization is greatly reduced (Figure 2e), suggesting that the strength of the EM field is related to the yield of the reaction.[39] This experiment can be rationalized as a power-dependence one.[26]

Our results clearly establish that site-selective extraction of hot carriers is the main consideration in determining the reactivity of the system, not hot carrier generation alone. While hot-carrier induced dissociation reactions, involving diatomic species such as $H_2$, $O_2$ and CO, had been recently described at the molecular level,[24, 26, 27, 43] we note that our reaction undergoes a six hot electron and a six protons ($H^+$) transfer to generate the final desired product (4-ATP) and two water molecules ($H_2O$) as side products.[30] A complex molecular mechanistic study is out of the scope of the presented work.[44] However, density functional theory (DFT) calculations for the first intermediate species show that proton transfer should be the limiting step with an estimated energy of $\Delta G \approx 0.35$ eV, compared to electron transfer that has essentially no activation barrier (see Figure S4). We do require the transfer of six individual electrons, that is, a single electron transfer event is not sufficient to report on the conversion. However, the applicability of our results is then broadened as the reaction can proceed with electrons of any energy, but only where they are generated in high enough quantities, as already described. The results should therefore more accurately reflect the locations where carriers are preferentially reaching the surface, thus highlighting locations of high reactivity in these nanoantennas (in a molecule or energy independent fashion). We note that the ability to locally tune the surface chemistry and reactivity of nanoantennas could open important avenues for reactive-spot and hot-spot modification.

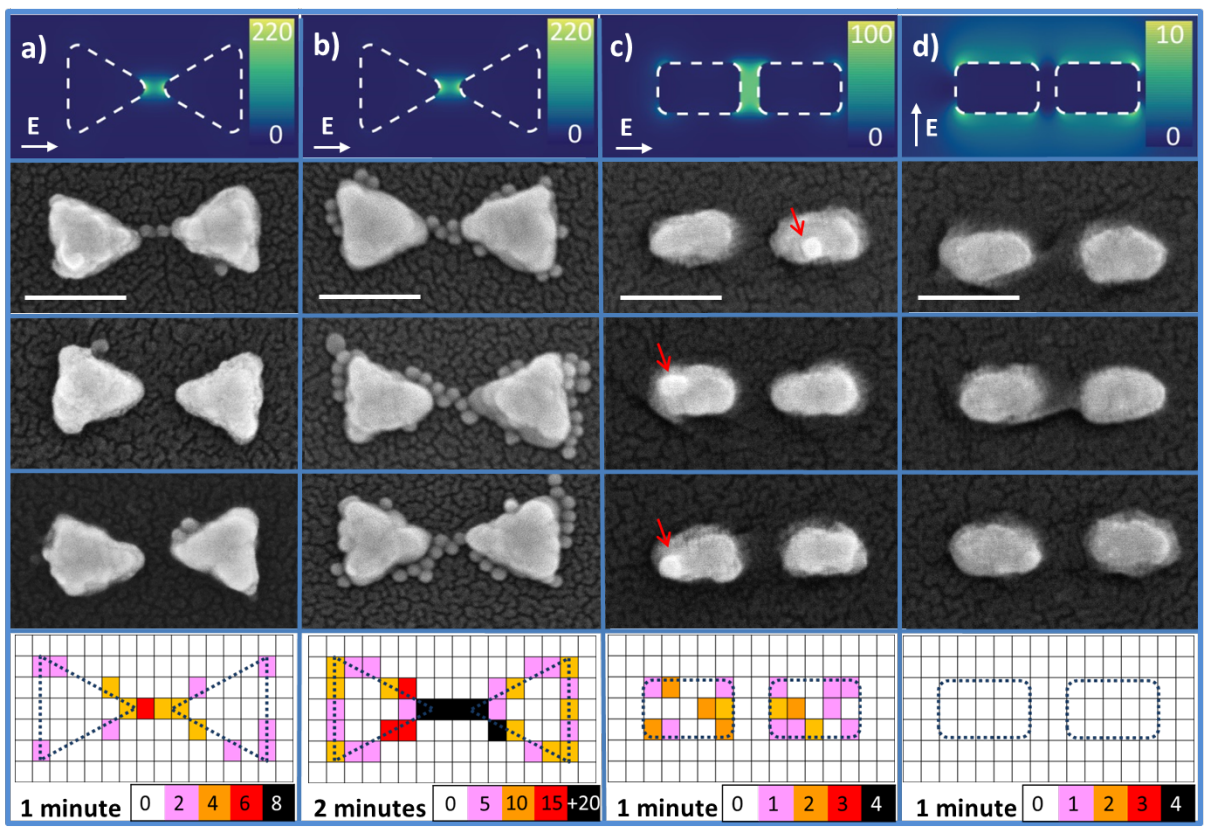

**Figure 5| Mapping hot-electron conversion in Ag bow-tie and Ag bar-dimer antennas for different illumination times and polarizations.** Au reporter particle binding to 4-NTP coated Ag bow-ties after **a)** 1 minute and **b)** 2 minutes of illumination with parallel polarization at 633 nm in 0.1 HCl. Au reporter particle binding to 4-NTP coated Ag bar-dimers conversion after 1 minute of **c)** parallel polarized illumination and **d)** perpendicular polarized illumination at 633 nm in 0.1 HCl. **a-d)** Top-panels show the near-field distribution calculated via FDTD simulations for each case at 633nm. Middle panels illustrate representative SEM images of the localized 15 nm AuNPs forming an amide bond with the converted molecules on the antenna (4-ATP). Scale bars: 100 nm. Bottom panels show the collapsed localizations over 100 antennas. Only antennas orientated between 0° and 20° respect to the incident polarization were taken into account (see Figure S3). Colour bar indicates the number (range) of particles localized.

Low efficiency in hot-carrier induced chemical reactions is the main unsolved problem that can pave the way in the field to large scale application of these concepts.[21, 45] Figures 4 and 5 show a strong spatial-energy dependence of the generated carriers and their extraction, both from first principle calculations and experiments. Reactions requiring highly energetic carriers will proceed only in a very small fraction of the antenna's surface. Also reactions involving high density of electrons will be strongly localized. Thus these results can open new avenues for the design of much more efficient nanoscale plasmonic systems for hot-carrier transport-driven chemical reactions.[29, 46] Our results on the transport and nanoscale localization of extracted carriers show that by tuning the strength of the EM field within the plasmonic antenna and by reducing the mean free-path of the carriers with high-local curvature tips we expect dramatic enhancement in the efficiency of hot-carrier induced chemical reactions and/or Schottky barrier photodetection and energy conversion approaches.

**Nanoscale surface chemistry: accessing reactive spots and EM hot spots**

As noted previously, the method used here to map the reactivity of plasmonic antennas relies on the ability to locally change the surface chemical properties of the antennas by selective hot-electron mediated reduction of the molecular layer (Figure 1). The ability to have highly localized differential chemical reactivity on the nanoantenna's surface can be further used to guide molecules, proteins and other nanomaterials (i.e. catalytic NPs) to specific regions of the nanoantenna.[47] For instance, small catalytic particles, which by themselves do not have a significant far-field cross section, can be incorporated to larger size structures that possess an large electric dipole moments to perform as nanoantennas.[48]

Furthermore, through careful design of antennas with sharp tips present only at the structure's hot-spots, this method could serve as an efficient, self-guided way of modifying only high EM field regions whilst keeping most of the antenna chemically passive (i.e. not reactive). In this way we can access both the most reactive spots (in terms of hot electron transfer) and the most plasmonically active regions (hot spots) of the nanoantennas. Attaching molecules to regions with the highest hot electrons extraction and/or the highest EM fields can further improve the efficiency of such photon or electron driven processes as photocatalysis, (bio)sensing, energy conversion, and imaging. This easy, fast and cheap strategy could serve as a selective and large-scale method of positioning molecules/nanomaterials in a variety of plasmonic nanoantenna reactive spots or hot spots. Photo-polymerization[49], three-photon absorption of disulphuric species[50] or local solvent heating strategy[51], among others, have been recently employed for high EM field hot spots modification. Here we have demonstrated that reactive spots can also be selectively accessed (Figure S5). We have shown this proof of concept by using 4-NTP coated Ag antennas and 15 nm AuNPs capped with a carboxylic-acid terminated molecular layer though, in principle, any amino reactive reaction may be used. By completing the reaction with a reducing agent like sodium borohydride ($NaBH_4$), this method should also be useful on Au nanoantennas, where selective hot-electrons reduction from 4-NTP to 4,4'-dimercaptoazobenzene (4,4'-DMAB) takes place.[52] Electrostatic-guided interaction can also be exploited due inherent differences between amino and nitro terminal groups.

## Conclusions

We have shown the ability to map a hot-electron reduction reaction on Ag nanoantennas with 15 nm spatial resolution and corroborated the spatially highly confined surface chemistry by first-principles calculations of hot carrier generation and transport. Our results progressively traced the reactivity in plasmonic antennas, highlighting strong dependence of the reactivity on the electromagnetic field distribution within the metal. Our theoretical treatment of plasmonic hot carrier generation and transport confirm nanoscale localization of high-density carrier regions required to drive this multi-electron chemical reaction, and predict an inverse relation between collected carrier density and transport distance. Polarization resolved experiments demonstrate that the high EM field intensities and the absorption of the nanoantennas are necessary to drive the reaction. Improved design of highly reactive and efficient antennas should benefit from these results. As we have shown, local surface chemistry can be tuned by employing this method, opening new possibilities for accessing regions of highly concentrated photon and electron densities. Positioning of nanomaterials or molecules in these regions is now possible and should boost research and applications in plasmonic hot-carrier science.

**Data availability.** The data that support the findings of this study are available from the corresponding authors on request (datainquiryEXSS@imperial.ac.uk).

## Acknowledgements

E.C. and J.C. are supported by Marie Curie fellowships of the European Commission. S.A.M. acknowledges the EPSRC Reactive Plasmonics project EP/M013812/1, the Office of Naval Research, the Royal Society, and the Lee-Lucas Chair in Physics. P.N. and S.A.M. acknowledge financial support from NG NEXT for this project. W.X. and S.S. acknowledge financial support from UDE and CENIDE. A.S.J. acknowledges financial support from a Goldwater Scholarship and a Marshall Scholarship. We want to thank Alberto Lauri and Bernd Walkenfort for helping us in the fabrication of some Ag control samples and SEM imaging, respectively. We thank Dr. Giulia Tagliabue, Professor William A. Goddard, III and Professor Harry A. Atwater for helpful discussions.

## Additional information

**Supplementary Information** accompanies this paper.
**Competing financial interests:** The authors declare no competing financial interests.